\newcommand{\CLASSINPUTbottomtextmargin}{1.75cm}
\newcommand{\CLASSINPUToutersidemargin}{1.7cm}

\documentclass[12pt,draftclsnofoot,onecolumn]{IEEEtran}

\usepackage{amsmath, mathrsfs, mathtools}
\usepackage{amsfonts}
\usepackage{epstopdf}
\usepackage{amssymb}
\usepackage{amsthm}
\usepackage{dsfont}
\usepackage{color}
\usepackage{graphicx}
\usepackage{caption}
\usepackage{subcaption}
\usepackage{algorithm}
\usepackage{algpseudocode}
\usepackage{acro}

\usepackage[backend=bibtex,sorting=none]{biblatex}
\usepackage{balance}
\let\labelindent\relax
\usepackage{enumitem}

\newcommand{\subparagraph}{}
\usepackage[compact]{titlesec}
\titlespacing*{\section}{15pt}{1.2\baselineskip}{0.9\baselineskip}

\theoremstyle{remark}
\newtheorem{theorem}{\bf Theorem}

\long\def\comment#1{}

\newcommand{\ben}{\begin{enumerate}}
\newcommand{\een}{\end{enumerate}}

\newcommand{\beq}{\begin{equation}}
\newcommand{\eeq}{\end{equation}}

\newcommand{\bi}{\begin{itemize}}
\newcommand{\ei}{\end{itemize}}

\DeclareMathOperator*{\argmax}{arg\,max}

\newcommand{\CC}{\mathbb{C}}

\newcommand{\RR}{\mathbb{R}}

\newcommand{\EE}{\mathbb{E}}

\newcommand{\cv}{{\bf c}}

\newcommand{\ev}{{\bf e}}

\newcommand{\hv}{{\bf h}}

\newcommand{\rv}{{\bf r}}
\newcommand{\sv}{{\bf s}}

\newcommand{\yv}{{\bf y}}
\newcommand{\zv}{{\bf z}}

\newcommand{\Hm}{{\bf H}}
\newcommand{\Id}{{\bf I}}

\newcommand{\Rm}{{\bf R}}

\newcommand{\Xm}{{\bf X}}
\newcommand{\Ym}{{\bf Y}}
\newcommand{\Zm}{{\bf Z}}

\newcommand{\tauv}{\hbox{\boldmath$\tau$}}

\renewcommand{\det}{{\hbox{det}}}

\newcommand{\SNR}{{\sf SNR}}

\renewcommand{\Re}{{\rm Re}}

\newcommand{\herm}{{\sf H}}

\DeclareAcronym{bs}{
	short = BS,
	long  = base station
}
\DeclareAcronym{iot}{
	short = IoT,
	long  = Internet-of-Things
}
\DeclareAcronym{mimo}{
	short = MIMO,
	long  =  multiple-input multiple-output
}
\DeclareAcronym{mu}{
	short = MU,
	long  =  multi-user
}
\DeclareAcronym{mrc}{
	short = MRC,
	long  = maximum ratio combining
}
\DeclareAcronym{ura}{
	short = U-RA,
	long  = unsourced random access
}
\DeclareAcronym{ad}{
	short = AD,
	long  = activity detection
}
\DeclareAcronym{ofdm}{
	short = OFDM,
	long  = orthogonal frequency-division multiplexing
}
\DeclareAcronym{tbm}{
	short = TBM,
	long  =  tensor-based modulation
}
\DeclareAcronym{hygamp}{
	short = HyGAMP,
	long  = hybrid generalized approximate message passing
}
\DeclareAcronym{scl}{
	short = SCL,
	long  = successive cancellation list
}
\DeclareAcronym{sic}{
	short = SIC,
	long  = successive interference cancellation
}
\DeclareAcronym{ml}{
	short = ML,
	long  = maximum likelihood
}
\DeclareAcronym{lsfc}{
	short = LSFC,
	long  = large-scale fading coefficient
}
\DeclareAcronym{mmvamp}{
	short = MMV-AMP,
	long  = multiple measurement vector approximate message passing
}
\DeclareAcronym{sinr}{
	short = SINR,
	long  = signal-to-interference-and-noise ratio
}
\DeclareAcronym{rss}{
	short = RSS,
	long  = received signal strengths
}
\DeclareAcronym{dl}{
	short = DL,
	long  = downlink
}
\DeclareAcronym{mmtc}{
	short = mMTC,
	long  = massive machine-type communication
}
\DeclareAcronym{noma}{
	short = NOMA,
	long  = non-orthogonal multiple-access
}
\DeclareAcronym{pme}{
	short = PME,
	long  = posterior mean estimator
}
\DeclareAcronym{amp}{
	short = AMP,
	long  = approximate message passing
}
\DeclareAcronym{pupes}{
	short = PUPEs,
	long  = per-user probabilities of error
}
\DeclareAcronym{npc}{
	short = NPC,
	long  = No power control
}
\DeclareAcronym{sci}{
	short = SCI,
	long  = Statistical channel inversion
}
\DeclareAcronym{lmmse}{
	short = LMMSE,
	long  = linear minimum mean squared error
}
\DeclareAcronym{crc}{
	short = CRC,
	long  = cyclic redundancy check
}
\DeclareAcronym{rv}{
	short = RV,
	long  = random variables
}
\DeclareAcronym{iid}{
	short = i.i.d.,
	long  = independent and identically distributed
}
\DeclareAcronym{DFT}{
	short = DFT,
	long  = discrete Fourier transform
}

\usepackage{vmr-symbols-vecbold}

\DeclareSymbolFontAlphabet{\amsmathbb}{AMSb}%

\newcommand{\lro}[1]{\lefto({#1}\right)}																
\newcommand{\lrbo}[1]{\lefto \lbrace {#1} \right \rbrace}															
\newcommand{\lrho}[1]{\lefto [ {#1} \right ]}																				

\newcommand{\lr}[1]{\left({#1}\right)}																
\newcommand{\lrb}[1]{\left \lbrace {#1} \right \rbrace}															

\safemath{\dopplerspread}{B_D}																								
\safemath{\delayspread}{T_D}																									
\safemath{\nc}{n\sub{c}}																										
\safemath{\nf}{n\sub{f}}																										
\safemath{\efa}{p\sub{sc}}
\safemath{\efb}{p\sub{cs}}
\safemath{\ef}{\epsilon\sub{f}	}
\safemath{\nd}{n\sub{d}}																										
\safemath{\ntx}{n\sub{t}} 																											
\safemath{\nrx}{n\sub{r}}																											
\safemath{\ntxt}{\tilde{n\sub{t}}}																											
\safemath{\cb}{\ensuremath{L}} 																								
\safemath{\cl}{\ensuremath{n}} 																								
\safemath{\txanto}{{\ensuremath{\tilde{m}_t}}} 																		
\safemath{\cs}{M} 																														
\safemath{\idPustm}{\ensuremath{S_{k}}}
\safemath{\error}{\ensuremath{\epsilon}} 																				
\safemath{\eexp}{\ensuremath{\mathcal{E}}} 																			
\safemath{\nsubc}{n\sub{s}}			 																						
\safemath{\nofdm}{n\sub{o}} 																									
\safemath{\bc}{\ensuremath{B_c}} 																							
\safemath{\ts}{\ensuremath{T_s}} 																							
\safemath{\nrb}{\ensuremath{n_{rb}}} 																						
\safemath{\rul}{\ensuremath{\rho\sub{ul}}}
\safemath{\rdl}{\ensuremath{\rho\sub{dl}}}

\safemath{\nres}{\ell}
\safemath{\nr}{n\sub{r}}
\safemath{\maxk}{M^*\lr{\nres, \nsubc, \nofdm, \epsilon, \rho}}
\safemath{\Rmax}{R^*}
\safemath{\Emin}{E\sub{b}^*/N_0}
\safemath{\Eminf}{\frac{E\sub{b}^*}{N_0}}
\safemath{\np}{\ensuremath{n\sub{p}}}
\safemath{\ndf}{\ensuremath{\bar{n}\sub{d}}}
\safemath{\npf}{\ensuremath{\bar{n}\sub{p}}}
\safemath{\code}{\ensuremath{\mathcal{C}}}
\safemath{\err}{\ensuremath{\epsilon}}
\safemath{\rp}{\ensuremath{\rho\sub{p}}}
\safemath{\rd}{\ensuremath{\rho\sub{d}}}
\safemath{\cohtime}{\ensuremath{T\sub{c}}}
\safemath{\cohbw}{\ensuremath{B\sub{c}}}
\safemath{\nmax}{\ensuremath{\ell\sub{m}}}
\safemath{\ntot}{\ensuremath{n\sub{tot}}}
\safemath{\nul}{\ensuremath{n\sub{ul}}}
\safemath{\ndl}{\ensuremath{n\sub{dl}}}

\safemath{\yp}{\ensuremath{\randvecy_{\nu}^{(\text{p})}}}
\safemath{\yd}{\ensuremath{\randvecy_{\nu}^{(\text{d})}}}
\safemath{\ypd}{\ensuremath{\vecy_{\nu}^{(\text{p})}}}
\safemath{\ydd}{\ensuremath{\vecy_{\nu}^{(\text{d})}}}

\safemath{\ypf}{\ensuremath{\bar{\randvecy}_{\nu}^{(\text{p})}}}
\safemath{\ydf}{\ensuremath{\bar{\randvecy}_{\nu}^{(\text{d})}}}
\safemath{\ypdf}{\ensuremath{\bar{\vecy}_{\nu}^{(\text{p})}}}
\safemath{\yddf}{\ensuremath{\bar{\vecy}_{\nu}^{(\text{d})}}}

\safemath{\xp}{\ensuremath{\vecx^{(\text{p})}}}
\safemath{\xd}{\ensuremath{\randvecx_{\nu}^{(\text{d})}}}
\safemath{\xdd}{\ensuremath{\vecx_{\nu}^{(\text{d})}}}

\safemath{\xpf}{\ensuremath{\bar{\vecx}^{(\text{p})}}}
\safemath{\xdf}{\ensuremath{\bar{\randvecx}_{\nu}^{(\text{d})}}}
\safemath{\xddf}{\ensuremath{\bar{\vecx}_{\nu}^{(\text{d})}}}

\safemath{\xdb}{\ensuremath{\overline{\randvecx}^{(\text{d})}}}
\safemath{\Pxd}{\ensuremath{P_{\randvecx^{(\text{d})}}}}

\safemath{\xpbar}{\ensuremath{\overline{\matX}^{(\text{p})}}}
\safemath{\xdbar}{\ensuremath{\overline{\randmatX}^{(\text{d})}}}

\safemath{\xdv}{\ensuremath{\randvecx^{(\text{d})}}}
\safemath{\xdbarv}{\ensuremath{\overline{\randvecx}^{(\text{d})}}}
\safemath{\ydv}{\ensuremath{\randvecy^{(\text{d})}}}

\safemath{\xdr}{\ensuremath{\matX^{(\text{d})}}}

\safemath{\ttx}{\ensuremath{\tau\sub{tx}}}
\safemath{\trx}{\ensuremath{\tau\sub{rx}}}
\safemath{\ack}{\ensuremath{\mathrm{s}}}
\safemath{\nack}{\ensuremath{\mathrm{c}}}

\newcommand{\prob}[1]{\ensuremath{\mathbb{P}\lrho{#1}}}

\safemath{\mI}{\ensuremath{i\lro{\randvecy ; \randvecx}}} 				

\safemath{\randveca}{\bm{A}}
\safemath{\randvecb}{\bm{B}}
\safemath{\randvecc}{\bm{C}}
\safemath{\randvecd}{\bm{D}}
\safemath{\randvece}{\bm{E}}
\safemath{\randvecf}{\bm{F}}
\safemath{\randvecg}{\bm{G}}
\safemath{\randvech}{\bm{H}}
\safemath{\randveci}{\bm{I}}
\safemath{\randvecj}{\bm{J}}
\safemath{\randveck}{\bm{K}}
\safemath{\randvecl}{\bm{L}}
\safemath{\randvecm}{\bm{M}}
\safemath{\randvecn}{\bm{N}}
\safemath{\randveco}{\bm{O}}
\safemath{\randvecp}{\bm{P}}
\safemath{\randvecq}{\bm{Q}}
\safemath{\randvecr}{\bm{R}}
\safemath{\randvecs}{\bm{S}}
\safemath{\randvect}{\bm{T}}
\safemath{\randvecu}{\bm{U}}
\safemath{\randvecv}{\bm{V}}
\safemath{\randvecw}{\bm{W}}
\safemath{\randvecx}{\bm{X}}
\safemath{\randvecy}{\bm{Y}}
\safemath{\randvecz}{\bm{Z}}
\safemath{\randvecphi}{\bm{\Phi}}

\safemath{\randmatA}{\amsmathbb{A}}
\safemath{\randmatB}{\amsmathbb{B}}
\safemath{\randmatC}{\amsmathbb{C}}
\safemath{\randmatD}{\amsmathbb{D}}
\safemath{\randmatE}{\amsmathbb{E}}
\safemath{\randmatF}{\amsmathbb{F}}
\safemath{\randmatG}{\amsmathbb{G}}
\safemath{\randmatH}{\amsmathbb{H}}
\safemath{\randmatI}{\amsmathbb{I}}
\safemath{\randmatJ}{\amsmathbb{J}}
\safemath{\randmatK}{\amsmathbb{K}}
\safemath{\randmatL}{\amsmathbb{L}}
\safemath{\randmatM}{\amsmathbb{M}}
\safemath{\randmatN}{\amsmathbb{N}}
\safemath{\randmatO}{\amsmathbb{O}}
\safemath{\randmatP}{\amsmathbb{P}}
\safemath{\randmatQ}{\amsmathbb{Q}}
\safemath{\randmatR}{\amsmathbb{R}}
\safemath{\randmatS}{\amsmathbb{S}}
\safemath{\randmatT}{\amsmathbb{T}}
\safemath{\randmatU}{\amsmathbb{U}}
\safemath{\randmatV}{\amsmathbb{V}}
\safemath{\randmatW}{\amsmathbb{W}}
\safemath{\randmatX}{\amsmathbb{X}}
\safemath{\randmatY}{\amsmathbb{Y}}
\safemath{\randmatZ}{\amsmathbb{Z}}
\safemath{\randmatSigma}{\mathbb{\Sigma}}
\safemath{\randmatPhi}{\mathbb{\Phi}}
\safemath{\randmatLambda}{\mathbb{\Lambda}}

\safemath{\matSigma}{\bm{\Sigma}}
\safemath{\matPhi}{\bm{\Phi}}
\safemath{\matLambda}{\bm{\Lambda}}

\def\Htran{\mbox{\tiny $\mathrm{H}$}}
\def\Ttran{\mbox{\tiny $\mathrm{T}$}}

\def\bphiu{\boldsymbol{\phi}} 

\usepackage{standard-macros}

\title{Coded Orthogonal Modulation for the Multi-Antenna Multiple-Access Channel}
\author{Alexander Fengler, Alejandro Lancho, Yury Polyanskiy
\IEEEcompsocitemizethanks{
\IEEEcompsocthanksitem The authors are with the Massachusetts Institute of Technology.
(Email: \{fengler,lancho,yp\}@mit.edu)
\IEEEcompsocthanksitem Research was sponsored by the United States Air Force Research Laboratory and the United States Air Force Artificial Intelligence Accelerator and was accomplished under Cooperative Agreement Number FA8750-19-2-1000. The views and conclusions contained in this document are those of the authors and should not be interpreted as representing the official policies, either expressed or implied, of the United States Air Force or the U.S. Government. The U.S. Government is authorized to reproduce and distribute reprints for Government purposes notwithstanding any copyright notation herein. Alejandro Lancho has received funding from the European Union’s Horizon 2020 research and innovation programme under the Marie Sklodowska-Curie grant agreement No 101024432. Alexander Fengler was funded by the Deutsche Forschungsgemeinschaft (DFG, German Research Foundation) – Grant 471512611. This work is also supported by the National Science Foundation under Grant No CCF-2131115.%
}%
}

\begin{document}

\maketitle

\begin{abstract}

    This study focuses on (traditional and unsourced) multiple-access communication over a single transmit and multiple $(M)$ receive antennas. We assume full or partial channel state information (CSI) at the receiver. It is known that to fully achieve the fundamental limits (even asymptotically) the decoder needs to jointly estimate all user codewords, doing which directly is computationally infeasible. We propose a low-complexity solution, termed coded orthogonal modulation multiple-access (COMMA), in which users first encode their messages via a long (multi-user interference aware) outer code operating over a $q$-ary alphabet. These symbols are modulated onto $q$ orthogonal waveforms. At the decoder a multiple-measurement vector approximate message passing (MMV-AMP) algorithm estimates \textit{several} candidates (out of $q$) for each user, with the remaining uncertainty resolved by the single-user outer decoders.
    Numerically, we show that COMMA outperforms a standard solution based on linear multiuser detection (MUD) with Gaussian signaling. Theoretically, we derive
    bounds and scaling laws for $M$, the number of users $K_a$, $\SNR$, and $q$, allowing to quantify the trade-off between receive antennas and spectral efficiency. The orthogonal signaling scheme is applicable to unsourced random access and, with chirp sequences as basis, allows for low-complexity fast Fourier transform (FFT) based receivers that are resilient to frequency and timing offsets.
\end{abstract}



\section{Introduction}
We look at the problem of unsourced multiple access in the uplink where $K_a$ users simultaneously attempt to transmit a $B$-bit message by
sending one of $2^B$ codewords from a common codebook $\mathcal{C} \in [q]^{n \times 2^B}$ using $q$-ary symbols. The unsourced setup with a common codebook \cite{Pol2017} addresses a setting where the user activity is sparse and the message length is short ($B\sim 100$ bits).  
We assume a quasi-static fading adder channel with additive white Gaussian noise (AWGN), i.e., the channel vectors $\hv_k \in \CC^M$ are assumed to be constant during the whole transmission duration.
Throughout most of this paper we assume that the channel vectors are known, in which case the unsourced aspect of the transmission can be neglected. 
The unsourced aspect can be reintroduced by using a pool of pilot sequences for channel estimation of which users pick one at random \cite{Fen2022,Gka2022, Ahm2022a}.  The capacity of multiple-access with known channel vectors at the receiver is given by \cite{Gol2003,Mar2016}
 \beq
 C = \log|1+\SNR\cdot\Hm\Hm'|
 \eeq 
 where $\Hm \in \CC^{M\times K}$ contains the channel vectors and $\SNR$ is the received signal-to-noise ratio.
Achieving this capacity requires joint-decoding of all user signals, which is often computationally infeasible. The most popular low-complexity solution is to employ 2-dimensional coded modulation at the transmitter, e.g. quadrature-phase-shift-keying (QPSK) or quadrature-amplitude-modulation (QAM), and use linear multi-user detection (MUD) in combination with single-user decoding, and potentially interference cancellation, at the receiver. In this work we study the alternative approach of orthogonal modulation schemes. In orthogonal modulation, a $q$-ary symbol is modulated into a vector of size $q$ by transmitting one of $q$ unit basis vectors of $\CC^q$.
This type of modulation has been applied in a wide variety of communication schemes, including unsourced multiple access (UMAC) schemes.
In UMAC schemes it was used most notably, in combination with compressed sensing in \cite{Ama2020a, Fen2021f,And2022a} but also for activity detection \cite{Ahm2022a}. 
Orthogonal signaling is also part of the LoRa physical layer \cite{Sem2019} in the form of chirp modulation.\footnote{Note that the LoRa sequences are only orthogonal in discrete time when sampled correctly \cite{Ben2022}.} Its low signal processing complexity at transmitter and receiver makes orthogonal chirp modulation very appealing for massive machine-type communication. In its current version LoRaWAN, the network protocol on top of LoRa, handles multiple access by means of simple collision resolution protocols (Aloha \cite{Abr1970}). Nonetheless, as demonstrated in, e.g., \cite{Ele2017}, it is in principle possible to allow for simultaneous transmission because the symbols from different users can be separated based on their channel vectors. 
In the simplest case of perfect synchronization the received signal can be written as 
\beq
	\Ym_i = \sum_{k=1}^{K_a} \sqrt{P}\hv_k \ev_{c_{k,i}}^T + \Zm_i , \quad i=1,...,n
	\label{eq:framesync}
\eeq 
where $K_a$ is the number of active users, $c_{k,i} \in [q]$ is the $i$-th symbol of the codeword of user $k$, $\ev_j, j=1,..,q$ represent some (column) unit basis vectors of $\CC^q$, and the entries of $\Zm\in \CC^{M\times q}$ are i.i.d. $\mathcal{CN}(0,1)$.
The signal $\Ym_i$ received at symbol slot $i$ is a matrix of dimensions $M\times q$, corresponding (for example) to different receive antennas and frequency bins, respectively.
It was recognized in \cite{Shy2020a}, in the context of unsourced random access, and in \cite{Ele2017}, in the context of LoRA MUD, that the users' messages can be separated by clustering the columns of $(\Ym_i)_{i=1,...,n}$ in the $q$ dimensional space. This approach becomes increasingly inefficient with a growing number of collisions, i.e., multiple users transmitting the same symbol in a symbol slot. 
In this paper we use the similarity of the detection problem to multiple-measurement vector compressed sensing \cite{Jin2013} and sparse regression codes (SPARC) \cite{Jos2012a,Rus2017} and propose an alternative detection algorithm based on multiple measurement vector approximate message passing (MMV-AMP) \cite{Kim2011}.\\

In the extreme case where $\hv_k = 1$ for every user $k$, users cannot be separated at all based on their channel. Nonetheless, it is still possible to recover the transmitted messages if they have been encoded by a forward-error correcting multiple-access code. The mentioned case, $\hv_k = 1$, reduces to the so-called unsourced A-channel \cite{Lan2022} by applying some sort of active set detection, e.g., thresholding.
 In this case, the decoders task is to recreate the list of transmitted message indices $j_1,...,j_{K_a}$ from the active sets
 $\mathcal{A}_i = \{j: y_{i,j} > \theta\}, i = 1,...,n$. 

 In this paper we propose a coding scheme called coded orthogonal modulation multiple-access (COMMA) which works as follows. At the transmitter side, the transmitted pulse sequences are chosen from a common A-channel code. At the receiver side, after channel estimation, the presented MMV-AMP algorithm is used for slot-wise detection of the users' symbols. The detector deliberately outputs more than one estimated symbol per user and an outer A-channel decoder is used to prune the wrong symbols. 
 As we will demonstrate in this paper, the achievable sum-spectral efficiency can significantly increase by concatenation with an outer unsourced A-channel code in addition to channel-based user separation. The use of unsourced forward-error correction allows to decode users messages even when their channel vectors are too close, which is a common source of error in the regime of small to medium $M$. 
 Numerical simulations show that the COMMA scheme outperforms classical two dimensional (one complex dimension) coded modulation schemes with linear MUD, where the latter are evaluated by means of finite-blocklength random coding achievability bound. Complex Gaussian signaling serves here as a proxy for popular modulation schemes like QPSK or QAM. Under perfect channel knowledge the observed gain in spectral efficiency of COMMA is almost a factor of two.

 
While a precise analysis of the effect of channel estimation error on the performance of COMMA is beyond the scope of this paper we provide some numerical examples at the end of the paper. The results show that the achievable spectral efficiency are cut in half but remain better than comparable 2-dimensional coded modulation schemes. 

Finally, the COMMA scheme can be made fully compliant with the unsourced random access scenario by introducing a pool of pilots and letting users pick one of them at random. This basic idea has been introduced in \cite{Fen2022} and been used in several of the best known UMAC schemes in the MIMO setting \cite{Gka2022, Ahm2022a}.
It was shown in \cite{Jia2023} that orthogonal modulation with the HyGAMP detector and without outer coding can achieve state-of-the-art UMAC performance.
Another closely
related scheme also using orthogonal modulation but in conjunction with compressed sensing was presented in \cite{Ago2023}. In \cite{Ago2023} bilinear generalized AMP \cite{Par2014} was used to jointly estimate data symbols and channel vectors without the use of pilots. This also showed state-of-the-art performance but at the expense of a higher receiver complexity. 
A detailed comparison is deferred to future work. Besides, we observe that uncoded orthogonal modulation is the most energy efficient in terms of $E_b/N_0$.  
Nonetheless, another practically relevant scenario is the case of fixed energy-per-symbol. Indeed, since the maximal transmit power is usually fixed by regulations, especially in unlicensed bands, it is not possible to further increase the power to allow for more users once a certain limit is reached. In this scenario, the concatenation of orthogonal signaling with an outer code allows to increase the user density and thus, the sum spectral efficiency.

\section{Information theory for a single-antenna A-channel}
Consider the problem where $K_a$ users transmit symbols from a $q$-ary input alphabet $[q] =\{1,\ldots,q\}$ over a channel that outputs the set of different transmitted symbols. Specifically, let $c_{i,j} \in [q]$ be the transmitted symbol from user $k \in [K_a]$ at channel use $i$. The channel output $y_i$ at channel use $i$ is given by
\begin{equation}\label{eq:channel1}
    y_i = \bigcup_{j=1}^{K_a}c_{i,k}.
\end{equation} 
We refer to this channel as the A-channel~\cite{Chang81,Bassalygo2000}. Note that here the receiver observes the set of transmitted symbols but not who transmitted them, and also not the multiplicity. 
 The A-channel was introduced by Chang and Wolf in~\cite{Chang81} as the ``$T$-user $M$-frequency channel without intensity information''. See \cite{Lan2022} for further details. 
\subsection{Unsourced A-channel with false alarms}
We now consider a noisy channel version of the model introduced in~\eqref{eq:channel1}, where at each channel use, the non-transmitted symbols are included in the output with probability $p\sub{fa}$. Hence, noise here refers to ``insertions''. The input-output relation is given by
\begin{equation}\label{eq:channel}
    \vecy = \lrbo{\bigcup_{k\in[K_a]}\vecc_{k}} \cup \vecz 
\end{equation}
where $\vecc_k\in [q]^n$ denotes the codeword transmitted by user $k$, and the length-$n$ vector $\vecz$ is such that $z_i$ contains each symbol in $[q]\backslash \{\bigcup_{k\in[K_a]}c_{k}\}$ with probability $p\sub{fa}$. Note that by setting $p\sub{fa} = 0$ we recover the classical noiseless A-channel. 
We next define the notion of UMAC code for the A-channel. Let $\binom{[a]}{b}$ denote the set of combinations of $b$-element subsets of $[a]$. Assume $q>K_a$, and let $W_k$, $k\in[K_a]$, denote the transmitted message by user $k$. An $(2^B,n,\epsilon)$-code for the unsourced A-channel~\eqref{eq:channel}, where $\vecc_k\in [q]^n$, consists of an encoder-decoder pair denoted as $f:[2^B]\mapsto [q]^n$ and $g:\lrbo{\bigcup_{k=1}^{K_a}{[q]\choose k}}^n\mapsto {[2^B] \choose K_a}$, respectively, 
satisfying the per-user probability of error (PUPE)
\begin{equation}\label{eq:pupe_def}
    P\sub{e}^{\rm (p)} \triangleq \prob{\{W_k \notin g(\vecy)\} \cup  \{W_k = W_{k'}, k\neq k'\}} \le \epsilon
\end{equation}
We assume that $\{W_k\}_{k=1}^{K_a}$ are independent and uniformly distributed on $[2^B]$, and that $f(W_k) = \vecc_k\in [q]^n$. For each type of error probability, we say the code achieves a \emph{rate} $R=B/n$.

Hence, we have $K_a$ users selecting randomly a codeword from a common codebook, and the decoder's task is to provide an estimate of the transmitted list of length $K_a$. Recall that in this paper, we assume $K_a$ is known at the receiver. The next theorem provides an achievability bound on the PUPE for the A-channel \eqref{eq:channel}.
\begin{theorem}
\label{thm:cover_pfa}
There exists a $(2^B,n,p_{\rm fa},\epsilon)$-code for the noisy unsourced $K_a$-user A-channel with PUPE satisfying $\epsilon \leq \epsilon\sub{A-ch}$, where
\begin{IEEEeqnarray}{lCl}
    \epsilon\sub{A-ch} &\triangleq& \left(\sum_{\ell=1}^{K_a-1}\frac{\ell}{K_a+\ell}\mathbb{E}\Biggl[\min\Biggl\{1,\binom{2^B-K_a}{\ell} \prod_{j=1}^{q}\left(\frac{j}{q}\right)^{\ell A_j(p_{\rm fa})}\Biggr\}\Biggr]\right. \nonumber\\
    && \qquad\qquad\qquad\qquad \left. {}+ \mathbb{E}\Biggl[\min\Biggl\{1,\binom{2^B-K_a}{K_a} \prod_{j=1}^{q}\left(\frac{j}{q}\right)^{K_a A_j(p_{\rm fa})}\Biggr\}\Biggr] + \frac{\binom{K_a}{2}}{2^B} \right). \label{eq:error_UB_ins}
\end{IEEEeqnarray}
 
Here, $A_j(p_{\rm fa})$ is the $j$-th element of a multinomial random vector with $n$ trials and $K_a$ possible outcomes with probabilities $\{p_j^{(\rm out)}(p_{\rm fa})\}_{j=1}^{q}$, where
\begin{IEEEeqnarray}{lCl}
    p_j^{(\rm out)}(p_{\rm fa}) &\triangleq& \mathbb{P}\left[|y| = j\right]\\
    &=& \sum_{\eta=1}^{\min{\{j,K_a\}}} \frac{p_\eta^{(\rm in)}}{Z_\eta} \binom{q-\eta}{j-\eta} p_{\rm fa}^{j-\eta}(1-p_{\rm fa})^{q-j} \label{eq:pk_noisy}
\end{IEEEeqnarray}
where $Z_\eta$ is a normalizing constant such that $\sum_{\eta=1}^{\min\{j,K_a\}} p_\eta^{(\rm in)} / Z_\eta = 1$. Furthermore,
\begin{equation}
    p_\eta^{(\rm in)} \triangleq \mathbb{P}\left[|\cup_{k=1}^{K_a}c_k|=\eta\right] = \frac{q! S(K_a,\eta)}{(q-\eta)!  q^{K_a}}. \label{eq:prob_card_in}
\end{equation}
where $S(\cdot,\cdot)$ denotes the Stirling number of the second kind \cite[Sec. 26.8.6]{olver10}. Finally, in~\eqref{eq:pk_noisy} we also used that
\begin{equation}
    \mathbb{P}\left[|z| = j-\eta \bigm| |\cup_{k=1}^{K_a}c_k| = \eta\right] = \binom{q-\eta}{j-\eta} p_{\rm fa}^{j-\eta}(1-p_{\rm fa})^{q-j}.
\end{equation}
\end{theorem}	
\begin{IEEEproof}
The proof follows along the lines of the proof of~\cite[Th.~1]{Lan2022}. 
\end{IEEEproof}

\section{Transmission Scheme}
We assume that $K_a$ users simultaneously attempt to transmit a $B$-bit message by
sending one of $2^B$ codewords of length $n$ from a common codebook $\mathcal{C} \in [q]^{n \times 2^B}$ with $q$-ary symbols.
We assume that the users employ orthogonal signaling. Let $\ev_q, q = 1,...,q$, denote the set of unit basis vectors of $\CC^q$.
The transmitted signal of user $k$ is
\beq
	\sv_k = [\ev_{c_{k,1}}^T, \ev_{c_{k,2}}^T,...,\ev_{c_{k,n}}^T] \in \{0,1\}^{nq}
\eeq 
where $\cv_k = [c_{k,1},...,c_{k,n}] \in \mathcal{C}$ is the codeword of user $k$, which we assume is uniformly chosen from $\mathcal{C}$.
We consider a SIMO quasi-static fading adder multiple-access channel 
\beq
	\yv[i] = \sum_{k=1}^{K_a} \sqrt{P}\sv_k[i]\hv_k^T + \zv[i] \quad i=1,...,nq \label{eq:fad_awgn_model}
\eeq 
with $\yv[i], \zv[i] \in \CC^M$. For the noise we assume $\zv[i] \sim \mathcal{CN}(0,\Id_M)$.
The energy-per-bit is then
$E_b/N_0 = \frac{nP}{B}$.
\section{Achievability}
\subsection{Non-fading case}
Here we evaluate explicitly the extreme case $M=1$ and $\hv_k=1$ for every user $k$. In this case users signals cannot be separated by means of their channel and we have to rely purely on multiple-access coding. 
We assume that the set of transmitted indices is estimated by thresholding.\footnote{An alternative employed in the literature is to take the $K_a + \Delta$ entries with the largest amplitudes. This approach can be shown to be equivalent to thresholding.} Let $\mathcal{A} = \{i: y[i]>\theta\}$. 
The choice of $\theta$ allows to trade-off between false-alarms and misdetections. Let $p_{\rm fa}$ denote the probability of false alarm, and $p_{\rm md}$ denote the probability of misdetection. Under AWGN noise (cf.~\eqref{eq:fad_awgn_model}), it follows that
\begin{IEEEeqnarray}{lCl}
    p\sub{fa} &=& Q(\theta) \label{eq:fa_awgn}\\
    p\sub{md} &=& 1 - \Ex{s}{\prob{z< \theta -\sqrt{P} - s\sqrt{P}}} \label{eq:md_awgn}
\end{IEEEeqnarray}
where $z\sim\setN(0,1)$ and $s\sim \text{Bino}(K_a-1,1/q)$. Using \eqref{eq:fa_awgn} and \eqref{eq:md_awgn} together with Theorem~\ref{thm:cover_pfa}, we have the following result.
\begin{theorem}\label{thm:cover_mdfa}
On the AWGN adder MAC, orthogonal signaling concatenated with a $K_a$-user $(2^B,n,p_{\rm fa},\epsilon)$-code for the unsourced A-channel and the threshold detection given by \eqref{eq:fa_awgn}--\eqref{eq:md_awgn}, achieves a PUPE $P_e$ satisfying
\beq
    P_e \leq n p\sub{md} + (1-n p\sub{md})\epsilon\sub{A-ch} \label{eq:PUPE_UB_awgn}
\eeq
where $\epsilon\sub{A-ch}$ is defined in~\eqref{eq:error_UB_ins}.
\begin{IEEEproof}
    By the union bound, the probability of misdetection in the received vector of length $n$ can be upper-bounded by $n p\sub{md}$, with $p\sub{md}$ given in \eqref{eq:md_awgn}. The result in~\eqref{eq:PUPE_UB_awgn} follows by assuming that any misdetection causes a full error, and using Theorem~\ref{thm:cover_pfa} with $p\sub{fa}$ given in \eqref{eq:fa_awgn}. 
\end{IEEEproof}
\end{theorem}
\subsubsection{Numerical examples} 
\begin{figure}[h]
\centering
\includegraphics[width=0.5\textwidth]{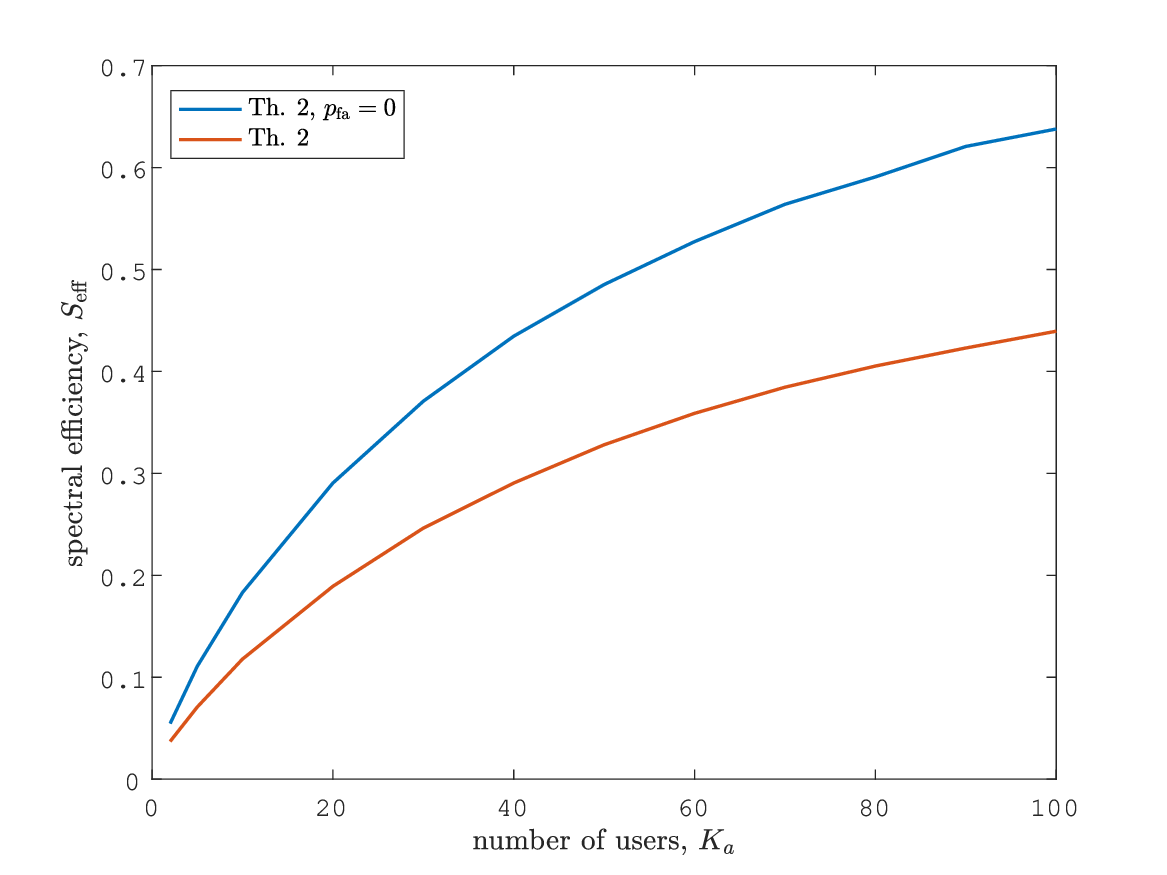}
\caption{Spectral efficiency $S\sub{eff} = KB/(nq)$ as a function of the number of active users $K_a$ given by optimizing the upper bound on the PUPE given in Theorem~\ref{thm:cover_mdfa} for $q = 2^8$, $n = 117$, $P=15~\mathrm{dB}$, $p\sub{md}=0.01/n$, and $\epsilon=0.05$.}
\label{fig:Seff_Ka}
\end{figure}

\begin{figure}[h]
\centering
    \includegraphics[width=.5\textwidth]{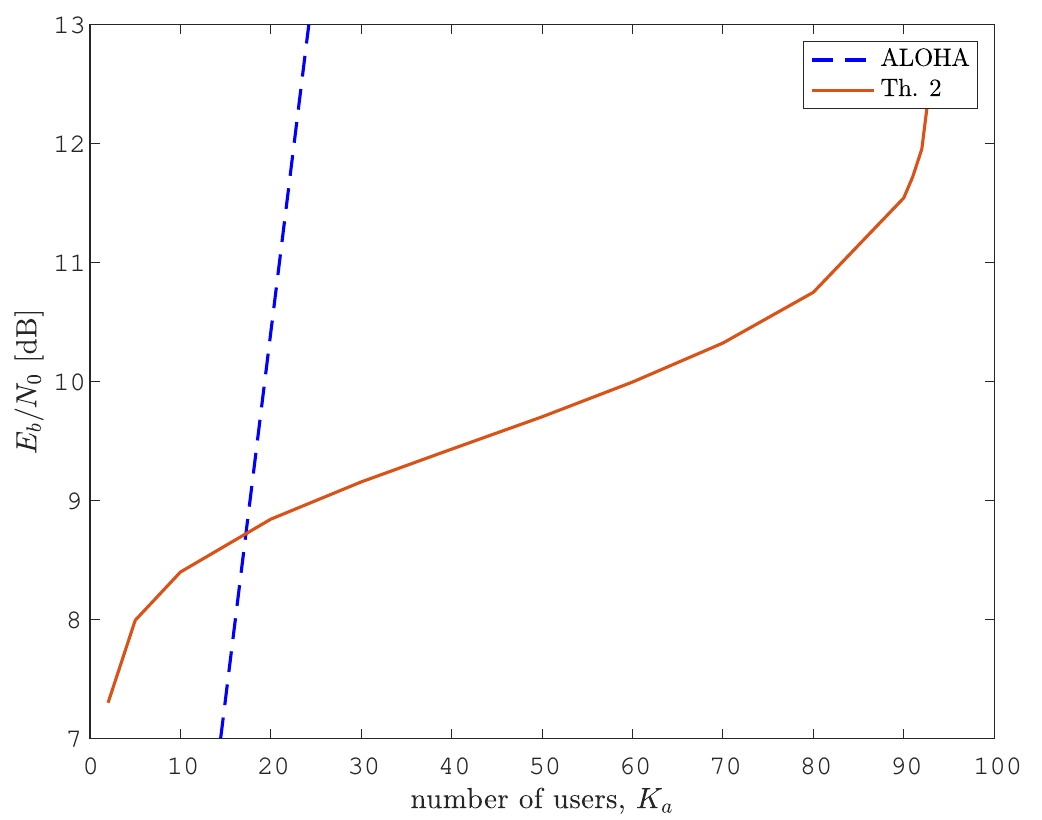}
\caption{$E_b/N_0$ in $\mathrm{dB}$ as a function of the number of active users $K_a$ given by optimizing the upper bound on the PUPE given in Theorem~\ref{thm:cover_mdfa} for $q = 2^8$, $B=200$, $n = 117$, $p\sub{md}=0.01/n$, and $\epsilon=0.05$.}
\label{fig:ebn0_Ka}
\end{figure}

In Fig.~\ref{fig:Seff_Ka}, we show the spectral efficiency given by $S\sub{eff} = KB/(nq)$ as a function of the number of active users $K_a$ obtained by computing the upper bound on the PUPE for the AWGN adder MAC with orthogonal signaling concatenated with a $K_a$-user unsourced A-chhanel code given by Theorem~\ref{thm:cover_mdfa}. We use $q = 2^8$, $n = 117$, $P=15~\mathrm{dB}$, $p\sub{md}=0.01/n$, and $\epsilon=0.05$. We plot a version of Theorem~\ref{thm:cover_mdfa} already derived in \cite[Th.~1]{Lan2022} where $p\sub{fa}$ is assumed to be zero (blue curve), and the new version introduced in this paper for general $p\sub{fa}$ (red curve). As we can observe, assuming $p\sub{fa}=0$ yields overly optimistic results. We further plot in Fig.~\ref{fig:ebn0_Ka} the energy per bit denoted by $E_b/N_0$ as a function of the number of active users again obtained by evaluating Theorem~\ref{thm:cover_mdfa}, for $q = 2^8$, $B=200$, $n = 117$, $p\sub{md}=0.01/n$, and $\epsilon=0.05$. We compare it against ALOHA, which was implemented by partitioning an $n_\text{tot} = nq$ long frame into $L$ subframes (we optimized $L$ for each $K_a$) and allowing users to randomly select a subframe. Successful decoding occurs only in the absence of collisions, and when a single-user decoder succeeds. Specifically, the error probability of the single-user decoder was obtained from the normal approximation for complex AWGN, where
\begin{IEEEeqnarray}{lCl}
    B \approx n_\text{tot}\log(1+P) - \sqrt{\frac{n_\text{tot}P(2+P)}{(1+P)^2}}Q^{-1}(\epsilon)
\end{IEEEeqnarray}
being $Q^{-1}(\cdot)$ the Gaussian $Q$-function. 
We can observe that the required $E_b/N_0$ grows slowly until $K_a \approx 90$ where the A-channel code is unable to correct the amount of introduced false alarms and keep the error probability $\epsilon\leq0.05$. Nonetheless, it outperforms ALOHA starting from $K_a \geq 17$.

\subsection{Fading Case}

\subsubsection{known $\hv_k$, Joint Detection}
In its most general form, the problem of recovering the set of transmitted symbols can be posed as a matrix recovery problem. Let $\Ym^i \in \CC^{M\times q}, i= 1,...,n$ be the matrix defined by $\Ym_i = [\yv[1 + (i-1)q],...,\yv[q + (i-1)q]]$. Furthermore, denote the matrix of channel vectors as $\Hm := [\hv_1,...,\hv_{K_a}]$. Then 
\beq
    \Ym_i = \Hm \Xm_i + \Zm_i   
    \label{eq:matrix_model}
\eeq
where $\Xm_i\in\{0,1\}^{K_a\times q}$ is defined by $\Xm_i = [\ev_{c_{1,i}}^T;...;\ev_{c_{K_a,i}}^T]$. We let the codeword symbols of each user $c_{k,i}$ be chosen from an appropriately sized common A-channel code.
Note that the $k$-th row of $\Xm_i$ contains the unit vector with a one at the transmitted symbol of user $k$ in slot $i$.
This formulation takes into account the constraint that each user transmits exactly one symbol per slot.
The maximum likelihood (ML) solution of \eqref{eq:matrix_model} is hard to analyze. Nonetheless, the resemblance of \eqref{eq:matrix_model} to an MMV recovery problem allows to solve it via MMV-AMP \cite{Kim2011,Liu2018a,Liu2018b,Fen2021c,Fen2022}. Specifically, 
\begin{align}
    \tau^{t+1}_j &= \frac{\|\Zm^t_{:,j}\|_2^2}{N} \quad j = 1,...,q,\\
    \tilde{\Xm}^{t+1} &= \underline{\eta}(\Hm^\herm \Zm^t + \Xm^t,\tauv_{t+1}),    
\label{eq:vamp_1} \\
\Xm^{t+1} &= \gamma \tilde{\Xm}^{t+1} + (1-\gamma)\tilde{\Xm}^{t}, \label{eq:vamp_damping}\\
    \Zm^{t+1}   &= \Ym_i - \Hm\Xm^{t+1} + \frac{K_a}{M} \Zm^{t}
    \langle \underline{\eta}^\prime(\Hm^\herm \Zm^{t} +
\Xm^{t},\tauv^{t+1})\rangle ,      \label{eq:vamp_2}
\end{align}
with $\Xm^0 = 0$ and $\Zm^0 = \Ym_i$. 
The additional damping step \eqref{eq:vamp_damping}, with some $\gamma \in (0,1]$, is necessary here to stabilize the algorithm in the regime where the dimension $q$ is comparable to the sample size $M$. See \cite{Fen2021c} for a discussion of the unstable behavior of MMV-AMP in the underdetermined regime. 
The {\em denosing} matrix-valued function
$\underline{\eta}: \CC^{K_a\times q}\times\RR^{q} \to \CC^{K_a\times q}$ with a
generic matrix argument $\Rm$  
is defined row-wise as
\begin{equation}
    \underline{\eta}(\Rm,\tauv) =
    \left [ \begin{array}{c} \eta_{1}(\Rm_{1,:},\tauv) \\ \vdots \\ 
\eta_{{K_a}}(\Rm_{K_a,:},\tauv) \end{array} \right ].  \label{etamatrix}
\end{equation}
Each row function $\eta_{k}: \CC^q\times\RR^q \to \RR^q$ is chosen as 
an estimate of the random vector $\ev \in \{\ev_1,...,\ev_q\}$
in the {\em decoupled} Gaussian observation model
\beq
\rv_k = \sqrt{P}\ev + \zv_k,
    \label{eq:vamp_decoupled}
\eeq
where $\zv_k$ is a complex Gaussian noise vector with components distributed as $
\mathcal{CN}(0,\text{diag}(\tauv))$. 
The choice of the specific denoising function will be discussed later. The term
$\langle \underline{\eta}^\prime(\Rm,\tauv)\rangle$ in \eqref{eq:vamp_2} 
is defined as
\beq
\langle \underline{\eta}^\prime(\Rm,\tauv) \rangle 
= \frac{1}{K_a}\sum_{k=1}^{K_a} \eta_{k}^\prime(\Rm_{k,:},\tauv), 
\eeq
where $\eta_{k}^\prime(\Rm_{k,:},\tauv) \in \CC^{q\times q}$ denotes
the matrix of the partial derivatives (Jacobian matrix) evaluated at the $k$-th
row of the matrix argument $\Rm$. 
For clarity, the $(a,b)$ element of the $q \times q$ matrix
$\eta_{k}^\prime(\Rm_{k,:},\tauv)$ is given 
by the partial derivative of the $a$-th component function of
$\eta_k$ with respect to the $b$-th component of the vector argument $\rv$,
i.e., 
\[ \left [ \eta_k^\prime (\rv, \tauv) \right ]_{a,b}  = \frac{\partial
[\eta_k(\rv, \tauv)]_a}{\partial [\rv]_b} \]
evaluated at $\rv = \Rm_{k,:}$. \\

Remarkably, in our problem $\eta$ as well as $\eta'$ can be written in closed form and are easy to evaluate. Specifically
\beq
    \eta_k(\rv,\tau) = \sqrt{P}\text{softmax}\left(\left(\frac{2\sqrt{P}\Re [r_a] - 1}{\tau_i^2}\right)_{a=1,...,q}\right).
\eeq 
and
\beq
    [\eta_k'(\rv,\tau)]_{ab} = \frac{\eta_k(\rv,\tau)_a}{\tau_a^2}\left[\sqrt{P}\delta_{ab}-\eta_k(\rv,\tau)_b\right]
\eeq
The equations are identical for all $k=1,...,K_a$. The function $\text{softmax}(\rv)$ is defined as
\beq
    \text{softmax}(\rv)_i = \frac{\exp(r_i)}{\sum_{j=1}^q \exp(r_j)}
\eeq 
After convergence, The output of the AMP algorithm $X_{k,i}$ can be interpreted as the marginal posterior probabilities that the $k$-th user transmitted symbol $i$. The typical approach in the closely related problem of SPARC decoding \cite{Rus2017} is to output the index with maximal posterior probability for each user.  We propose a modified version where the decoder, after convergence of AMP, outputs the $1+N_\text{fa}$ indices per-user with the largest posterior probabilities. These additional symbols can then be corrected by the outer A-channel code.

\begin{figure}[h]
\centering
\includegraphics[width=0.5\textwidth]{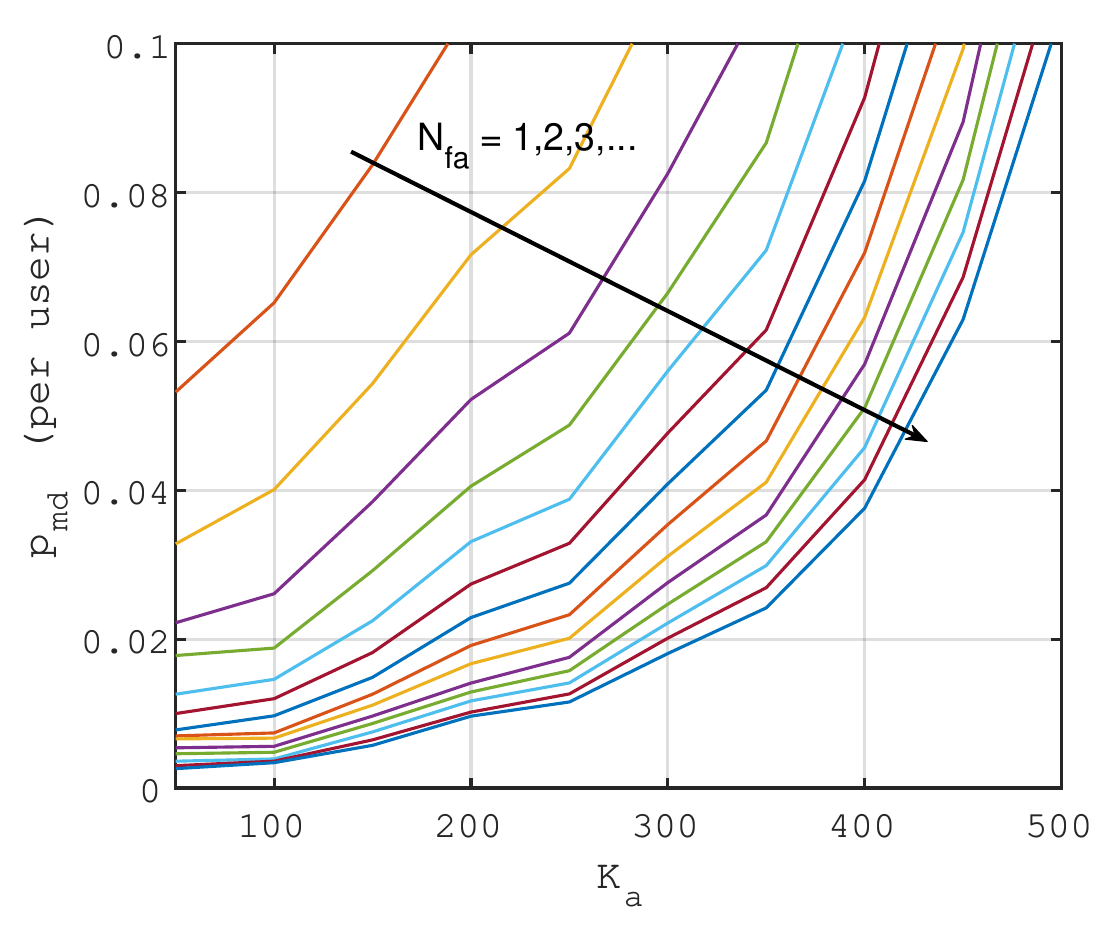}
\caption{Joint decoding with MMV-AMP and output of top $1+N_\text{fa}$ indices per-user + A-channel code. $q = 2^7, P=0.7, M=25, n=23$.}
\label{fig:pe_over_K_AMP}
\end{figure}

In Fig.~\ref{fig:pe_over_K_AMP} we simulate the probability per-user that at least one index of a transmitted sequence is missed if the decoder is allowed to output the $1+N_\text{fa}$ top indices per-user. We see that especially for small $p_\text{md}$ a few $N_\text{fa}$ can increase the number of active users significantly. Finally, if there is more than one message remaining in the output list of the A-channel code, the decoder outputs the one with the highest posterior-likelihood, i.e. $\hat{c}_k = \argmax_{c\in \mathcal{L}_k} \sum_{i=1}^n X_{k,c_i}$ where $\mathcal{L}_k$ is the list output of the outer decoder for user $k$. 

\subsubsection{Half-Space method/ Matched Filtering}
An alternative to ML, which allows for analysis over the whole parameter regime (including the overloaded case $K_a>q$), is the following assignment rule: For some fixed symbol slot $i$ let $\yv_j = \yv[j + (i-1)q], j=1,...,q$.
\begin{quotation}
{\bf Matched Filter (MF):} Symbol $j$ is added to the list of possibly transmitted symbols for user $k$ if $\text{Re}[\langle\yv_j,\hv_k\rangle] > \theta_k$, where
$\theta_k$ are chosen as parameters of the detector. A symbol can be assigned to the list of multiple users. 
\end{quotation}

\begin{theorem}
The MF detector concatenated with an outer single-user $(2^B,n,\left(1 + \frac{P/2}{P(K_a/q + K_a^{3/4}) + 1}\right)^{-M},\epsilon)$ A-channel code achieves a PUPE $P_e$ satisfying
\beq
    P_e \leq \exp\left(\log n - M\log \left[1+ \frac{P/2}{P(K_a/q + K_a^{3/4}) + 1}\right]\right) + \exp(-2\sqrt{K_a}) +\epsilon
\eeq 
\hfill $\square$
\end{theorem}
The above theorem shows that a given PUPE $P_e$ is achievable with
\beq
    M = \frac{K_a}{q} + \min\left\{K_a^{3/4},\sqrt{\frac{\log 1/P_e}{2}}K_a^{1/2}\right\} + 2/P + \log B - \log P_e
\eeq 
antennas and a sum spectral efficiency $S = K_a\log_2 q/q$ (under this scaling $p_\text{fa}\to 0$ and $R_\text{out}\to 1$). The second factor in the $\min$ arises from the alternative bound on the number of colliding users in a slot, given in the proof of the Theorem. This results quantifies how the choice of $q$ allows to trade-off between antennas and spectral efficiency.   

\begin{proof}
W.l.o.g., assume $k=1$ and that user $1$ transmits symbol $1$. Let $p_1(\yv_1)$ denote the distribution of $\yv_1 = \sqrt{P}\hv_1 + \sqrt{P}\sum_{k'\in S_1\setminus 1} \hv_k' \zv$ where $S_1$ denotes the (random) set of users that also transmit symbol $1$. Note that $|S_1|\sim\text{Bino}(K_a-1,1/q)$ and let $\theta_1$ be chosen such that $p_1(\Re[\langle\yv_1,\hv_1\rangle] > \theta_1) \geq 1 - p_\text{md}$. It then follows that for any event $A$
\beq
    p_1(A) = \sum_{s=0}^{K_a-1}p_1(A||S_1| = s)p(|S_1| = s)
    \label{eq:hspace_1}
\eeq 
and
\beq
\begin{split}
    p_1(\Re[\langle\yv_1,\hv_1\rangle] < \theta_1||S_1| = s) 
    &= P(\sqrt{P}\|\hv_1\|^2 + \sqrt{P}\Re[\langle\sum_{i=2}^{s+1}\hv_{i},\hv_1\rangle] + \Re[\langle \zv_1,\hv_1\rangle] < \theta_1) \\
    &= P(\sqrt{P}\|\hv_1\|^2 + \sqrt{(Ps + 1)/2}\|\hv_1\|\tilde{z} < \theta_1) \\
    &= Q\left(\frac{\sqrt{P}\|\hv_1\|^2-\theta_1}{\|\hv_1\|\sqrt{(Ps + 1)/2}}\right).
    \label{eq:hspace_2}
\end{split}
\eeq 
Hence, we can use \eqref{eq:hspace_1} and \eqref{eq:hspace_2} to compute $\theta_1$ numerically. \footnote{For the real model there is not factor $1/2$ in the denominator.}
Let $S'$ be the random set of users that transmit some symbol $q\neq 1$. Again, $|S'|\sim\text{Bino}(K_a-1,1/q)$. The expected number of false alarms for user 1 (averaged over $\hv_2,...,\hv_K,\zv_1,...,\zv_q$) can then be calculated as
\beq
    N_{\text{fa},1} = (q-1)\sum_{s=0}^{K_a-1} Q\left(\frac{\theta_1}{\|\hv_1\|\sqrt{(sP + 1)/2}}\right)p(|S'| = s). 
    \label{eq:nfa1}
\eeq 

We next use \eqref{eq:hspace_2} and \eqref{eq:nfa1} to derive an achievable scaling of the number of users. 
We choose $\theta_i = \alpha \sqrt{P} \|\hv_1\|^2$ for some $\alpha \in [0,1]$. It then follows that
\beq
    p_{\text{md},s} = p_1(\langle\yv_1,\hv_1\rangle < \theta_1||S| = s) \leq \exp\left(-\frac{2P(1-\alpha)^2\|\hv_1\|^2}{Ps + 1}\right) .
\eeq 
Taking the expected value over $\hv_1$ gives
\beq
    \EE[p_{\text{md},s}] \leq \left(1 + \frac{2P(1-\alpha)^2}{P(s-1) + 1}\right)^{-M}
\eeq 
A similar calculation for the probability of false alarms per section yields
\beq
    \EE[p_{\text{fa},s}] \leq \left(1 + \frac{2P\alpha^2}{Ps + 1}\right)^{-M}
\eeq 
where $\alpha$ can be chosen such that $p_\text{md}:=\sum_s p(s)\EE[p_{\text{md},s}] \leq \epsilon/(2n)$. Thus, an $(2^B,n,p_\text{fa},\epsilon/2)$ outer code will result in an average PUPE smaller than $\epsilon$. Both $p_{\text{fa},s}$ and $p_{\text{md},s}$ are increasing with $s$. Therefore we can bound them by the largest $s$. A basic tail bound on binomial RVs shows that $s\leq K_a/q + K_a^{3/4}$ holds with probability $1 - \exp(-2\sqrt{K_a})$. An alternative bounds is $s\leq K_a/q + \delta \sqrt{K_a}$ with probability $1-\exp(-2\delta^2)$.
Furthermore, we fix $\alpha = 1/2$ so
$p_\text{md} \leq \left(1 + \frac{P/2}{P(K_a/q + K_a^{3/4}) + 1}\right)^{-M} + \exp(-2\sqrt{K_a})$ and
$p_\text{fa} \leq \left(1 + \frac{P/2}{P(K_a/q + K_a^{3/4}) + 1}\right)^{-M}$.
Assuming that an outer $(2^B,n,p_\text{fa},\epsilon)$ A-channel code is used gives the result of the theorem.
\end{proof}

\section{MU-MIMO with coded modulation and linear multi-user detection}
\label{sec:MIMO}
\subsection{A MIMO finite-blocklength upper bound on the error probability}
We assume that each coded packet spans $n$ discrete-time channel uses.
For a given arbitrary transmission round, the received signal $\vecr[i] \in \mathbb{C}^{M} $ at the access point equipped with $M$ antennas at discrete-time $i\in[n]$ is modeled as 
  $\vecr[i] = \sum_{k=1}^{K_a}\vech_k q_k[i] + \vecz[i].$
Here, ${\vech}_{k}\in \mathbb{C}^{M}$ denotes the channel vector between user $k$ and the access point.
We use an uncorrelated Rayleigh-fading model where $\vech_{k} \sim \mathcal{CN}({\veczero}_{M},\mathbf{I}_{M})$ remains constant for the duration of a packet transmission. 
The vector $\vecz[i]\in \mathbb{C}^{M}$, with i.i.d. elements distributed as $\mathcal{CN}(0,1)$ models the additive noise.
Finally, $q_k[i]$ is the $i$-th symbol of the coded packet transmitted by user $k$.

We assume that the first $\np$ symbols in each coded packet are used to transmit pilot symbols used by the access point to estimate the channels, whereas the remaining $\nd=n-\np$ symbols contain the data. 
The $\np$-length pilot sequence of user $k$ is denoted by the vector $\bphiu_{k} \in \mathbb{C}^{\np}$.
It is designed so that $\| \bphiu_{k} \|^2  = \np$.
Users pick their pilot sequence from a set of $2^J$ unit-norm, random complex pilot sequences where each element phase is generated uniformly i.i.d. 
We use MMSE channel estimation~\cite[Sec.~3.2]{bjornson19}, for which the estimate $\widehat{\vech}_{k}$ of $\vech_k$ is given by
\begin{equation} 
 \widehat{\vech}_{k}  = \sqrt{P \np}{\bf Q}_{k}  \left({\bf V}^{\mathrm{pilot}} \bphiu_{k}\right).
\end{equation}
where
\begin{equation}\label{eq:covariance-matrix-pilot-signal}
  {\bf Q}_{k}  = \left( {\bf I}_{M}\left(\sum_{k^{\prime}=1}^{K_a} P \bphiu_{k^{\prime}}^{\Htran}\bphiu_{k} +  \sigma^2  \right)\right)^{-1} 
 \end{equation}
 and
 \begin{equation}\label{eq:vector-channel-UL-pilots}
   {\bf V}^{\mathrm{pilot}} = \sum_{k=1}^{K_a} \sqrt{P}{\vech}_{k}\bphiu_{k}^{\Ttran} +  {\bf Z}^{\mathrm{pilot}}.
   \end{equation}
 Here,  ${\bf Z}^{\mathrm{pilot}} \in \mathbb{C}^{M \times \np}$ is the additive noise with i.i.d. elements distributed as $\mathcal{CN}(0,1)$.

We assume that the access point uses the channel estimates $\{\widehat{\vech}_k\}_{k=1}^M$ to separate the users via linear combining. 
Hence, to recover the signal transmitted by user $k$, it projects the vector $\vecr[i]$, $i=n\sub{p}+1,\dots, n$, onto the MMSE linear combiner ${\vecu}_{k}$ given by  
\begingroup
\allowdisplaybreaks
 \begin{align}\label{eq:MMSE_combiner}
{\vecu}_{k} = \left(\sum\limits_{k^{\prime}=1}^{K_a}\widehat{\vech}_{k^{\prime}}\left(\widehat{\vech}_{k^{\prime}}\right)^{\Htran}  + {\bf Z}\right)^{-1}\widehat{\vech}_{k}
\end{align}%
\endgroup%
with ${\bf Z} =\sum\nolimits_{k^{\prime}=1}^{K_a} {\boldsymbol{\Phi}}_{k^{\prime}} + \frac{\sigma^{2}}{P}{\bf{I}}_M$, where $\boldsymbol{\Phi}_{k} = P\np {\bf Q}_{k}$. 
Let $v[i]=\vecu_k^{\Htran} \vecr[i]$ and $g=\vecu_k^{\Htran}\vech_k$.
Then, we can express $v[i]$ as $v[i]=gq_k[i]+z[i]$, where $z[i] = \sum_{k^\prime=1,k^\prime\ne k}^{K_a} {\vecu}_{k}^{\Htran}{\vech}_{k^\prime} q_{k^\prime}[i] +{\vecu}_{k}^{\Htran}{\vecz}[i]$ contains the additive noise and the residual interference from the other users after combining. 

To obtain an estimate of the packet error probability, we assume that the access point treats the channel estimate as perfect and the residual interference as noise by performing scaled nearest-neighbor decoding. 
A random coding analysis, performed under the assumptions that the data symbols $q_k[i]$ are i.i.d. and follow a $\mathcal{CN}(0,P)$ distribution, yields the following upper bound on the packet error probability $\epsilon$~\cite[Eq. (3)]{ostman20-09b} 
\begin{equation}
  \epsilon
  \leq \Pr\lrho{\sum_{i=\np+1}^n {\imath_s(q[i],v[i])} \leq\log\left(\frac{2^B-1}{U}\right)}. \label{eq:rcus_tail}
\end{equation}
Here $B$ is the number of transmitted information bits, $U$ is a uniformly distributed random variable on the interval $[0,1]$, and
\begin{equation}
\begin{split}
\imath_s(q[i],v[i]) &= -s \left|{v[i] - \widehat{g} q[i]}\right|^2 + \frac{s\abs{v[i]}^2}{1+sP\abs{\widehat{g}}^2} \\
&\quad + \log\lro{1+sP\abs{\widehat{g}}^2}\\
\end{split}
\end{equation}
with $\widehat{g}=\vecu_k^{\Htran}\widehat{\vech}_k$, and $s>0$ being an optimization parameter that can be used to tighten the bound. To efficiently evaluate the bound~\eqref{eq:rcus_tail}, we use the so-called saddlepoint approximation described in \cite[Section II.B]{ostman20-09b}.

\section{Numerical Results}
\subsection{Perfect CSI}
\begin{figure}[h]
\centering
\includegraphics[width=0.5\textwidth]{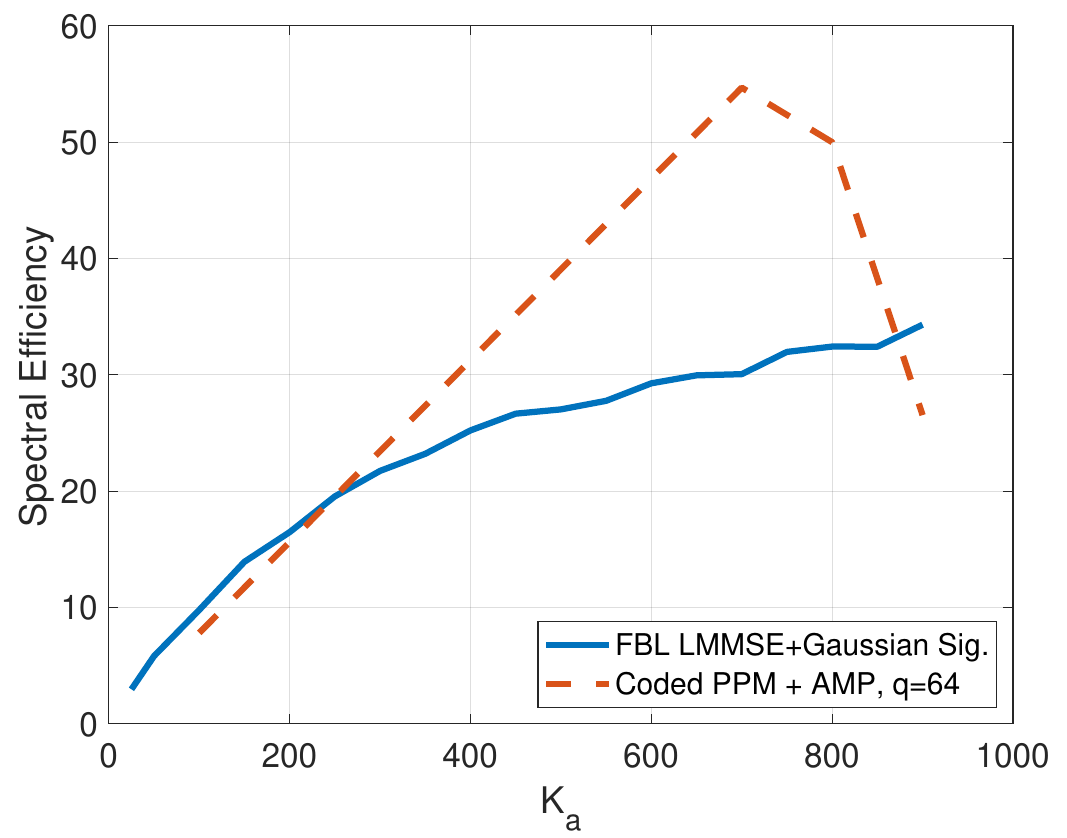}
\caption{Joint decoding with MMV-AMP and output of top $1+N_\text{fa}$ indices per-user. The $N_\text{fa}$ wrong indices are the corrected by an A-channel code. Shown is the sum spectral efficiency $K_aB/(n2^J)$. $Q = 2^6, P=0.3, M=50$.}
\label{fig:sumrate_over_K_AMP_M50}
\end{figure}

\begin{figure}[h]
\centering
\includegraphics[width=0.5\textwidth]{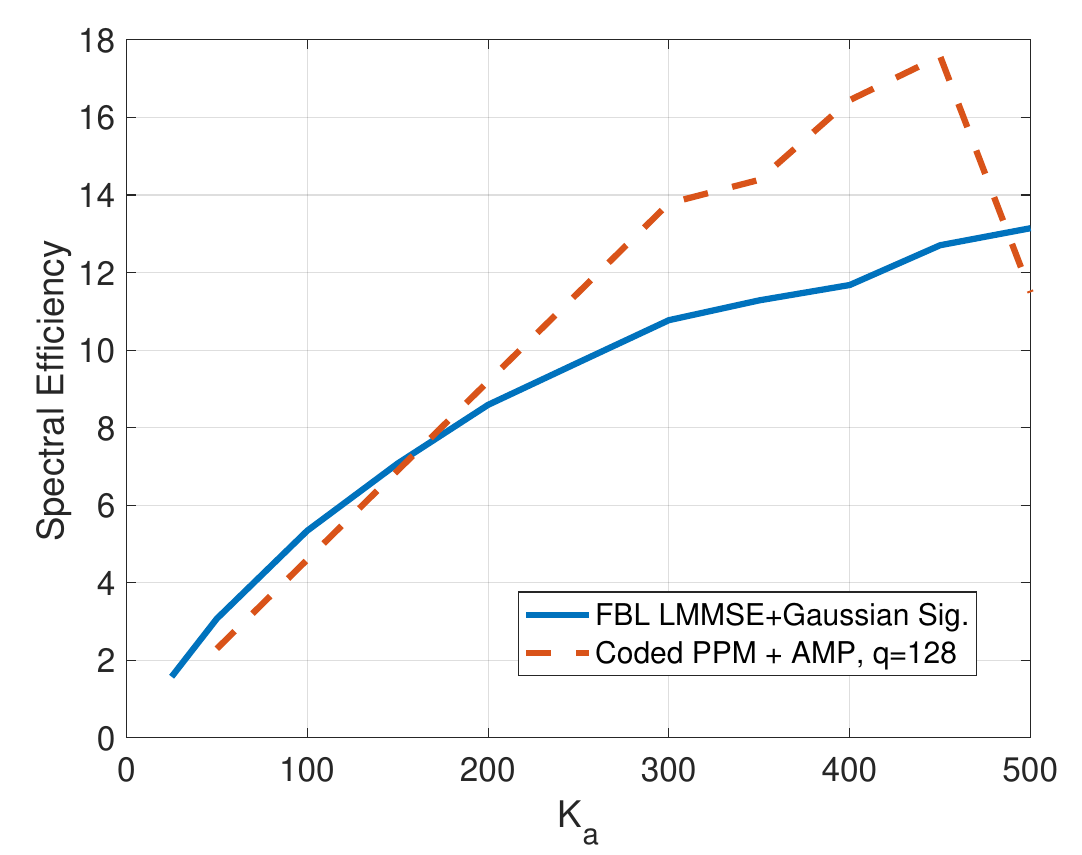}
\caption{Joint decoding with MMV-AMP and output of top $1+N_\text{fa}$ indices per-user. The $N_\text{fa}$ wrong indices are then corrected by an A-channel code. Plotted is the sum spectral efficiency $K_aB/(n2^J)$. $Q = 2^7, P=0.7, M=25$.}
\label{fig:sumrate_over_K_AMP_M25}
\end{figure}

Here we compare the results for orthogonal signaling with MMV-AMP user detection to the FBL bounds for classical MIMO coded modulation with linear MUD. For that, we fix the payload size to $B = 100$ bits and we fix the energy-per-symbol $P$. For the coded modulation scheme a symbol has duration $1/q$ compared to orthogonal signaling. The power is adjusted accordingly to $P/q$. Then we compute the minimum required blocklength $n_\text{tot}$ to achieve a PUPE $\leq 0.05$ and plot the sum spectral efficiency $S = K_a B/n_\text{tot}$. In Fig.~\ref{fig:sumrate_over_K_AMP_M50} and Fig.~\ref{fig:sumrate_over_K_AMP_M25} we assume perfect knowledge of the channel coefficients at the receiver. The FBL bounds are computed as described in Section~\ref{sec:MIMO}. For the COMMA curves we fix a symbol size $q$ and simulate the symbol detection errors of the MMV-AMP MUD algorithm for different values of $N_\text{fa}$. For each $N_\text{fa}$ we find a single-user A-channel code such that the sum of missed detection probability and decoding error probability is smaller than $0.05$, if such a code exist. The total blocklength is then given by $n_\text{tot} = n q$, where $n$ is the blocklength of the A-channel code that achieves the required error probability with the smallest $N_\text{fa}$. \footnote{Although we use Theorem~\ref{thm:cover_pfa} to find the required code rates, simulations show that the Tree Code from \cite{Ama2020a} is very close to the rates from Theorem~\ref{thm:cover_pfa}.}
The achievable spectral efficiencies almost double compared to the classical coded modulation schemes even under the strong assumption of Gaussian signaling. 
We can observe a linear growth of the spectral efficiency up to some point ($K=700$ in Fig.~\ref{fig:sumrate_over_K_AMP_M50} and $K=300$ in Fig.~\ref{fig:sumrate_over_K_AMP_M25}.)  after which the required number of $N_\text{fa}$ increases rapidly and lets the sum spectral efficiency go to zero. The change is more abrupt as $M$ grows large. This behavior can be confirmed by evaluations of the state evolution equations, which predict a jump of the missed detection rate from $0$ to $1$ at around $K/M = 90$ for the parameters of Fig.~\ref{fig:sumrate_over_K_AMP_M50}. Note, that the state evolution equations here are only valid in the limit $K,M \to \infty$. This confirms the intuition that outer coding improves the spectral efficiency only in the regime where $M$ is comparably small. For large $M$ high dimensional concentration guarantees that all channel vectors are concentrated in a thin shell. 

\subsection{Imperfect CSI}
\begin{figure}[h]
\centering
\includegraphics[width=0.5\textwidth]{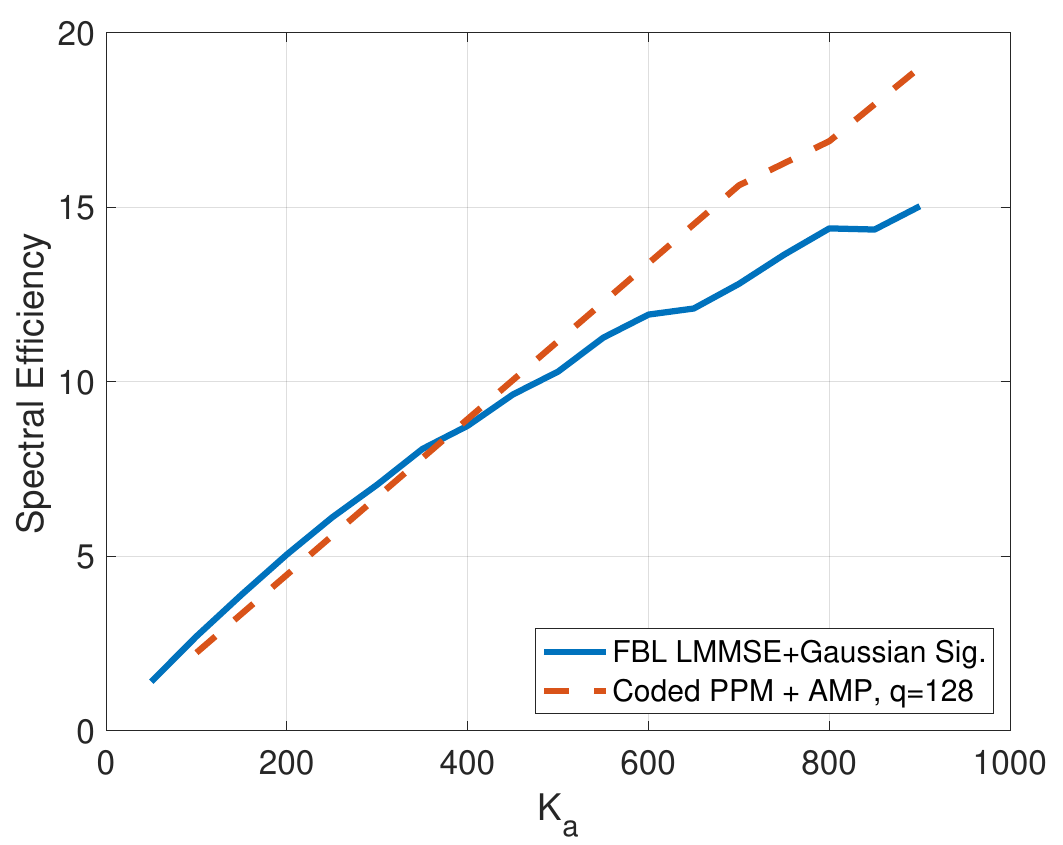}
\caption{Joint decoding with MMV-AMP and output of top $1+N_\text{fa}$ indices per-user. The $N_\text{fa}$ wrong indices are then corrected by an A-channel code. Plotted is the sum spectral efficiency $K_aB/(n2^J + L2^J)$ taking into account the transmission of $L$ $Q$-ary pilot symbols. $Q = 2^7, P=0.6, M=50, L=20$.}
\label{fig:sumrate_over_K_AMP_M50_cer}
\end{figure}

\begin{figure}[h]
\centering
\includegraphics[width=0.5\textwidth]{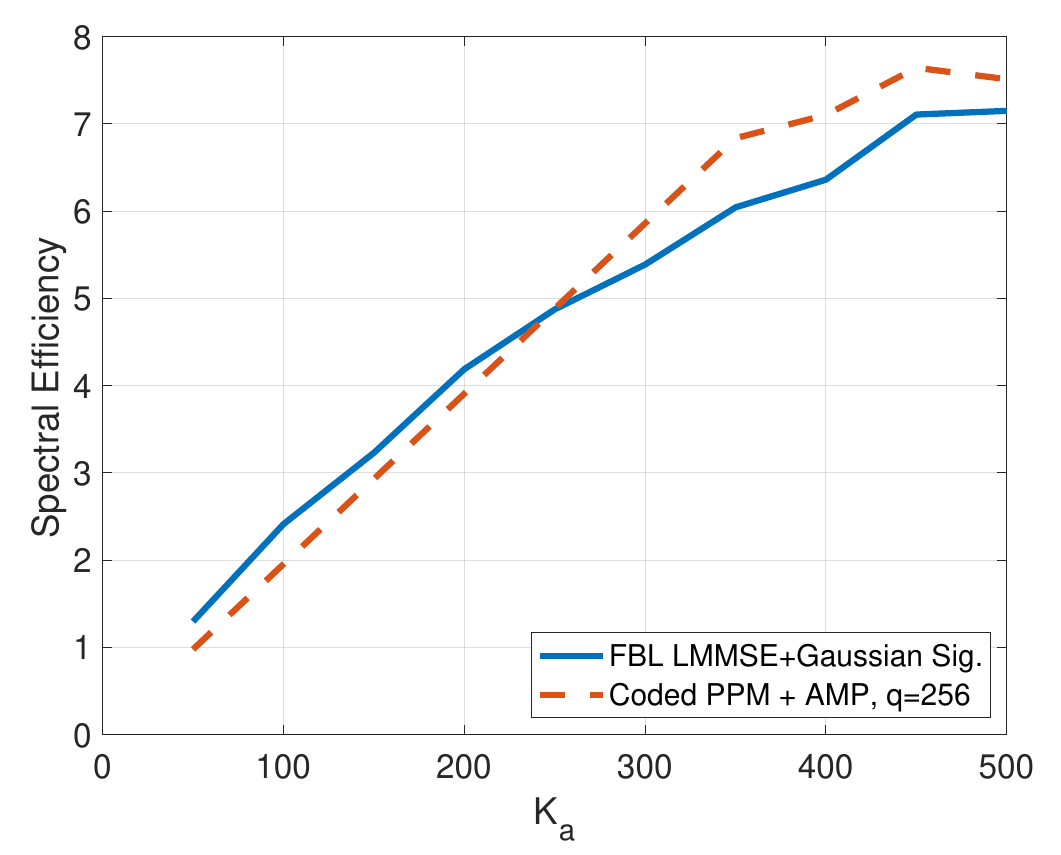}
\caption{Joint decoding with MMV-AMP and output of top $1+N_\text{fa}$ indices per-user. The $N_\text{fa}$ wrong indices are then corrected by an A-channel code. Plotted is the sum spectral efficiency $K_aB/(n2^J + L2^J)$ taking into account the transmission of $L$ $Q$-ary pilot symbols. $Q = 2^8, P=1.4, M=25, L=7$.}
\label{fig:sumrate_over_K_AMP_M25_cer}
\end{figure}

In Fig.~\ref{fig:sumrate_over_K_AMP_M50_cer} and Fig.~\ref{fig:sumrate_over_K_AMP_M25_cer} we plot the achievable spectral efficiencies as described in the previous section when the channels are first estimated by means of MMSE channel estimation over $n_\text{pilot}$ symbols.
The results show that the spectral efficiencies are almost cut in half compared to the case of perfect channel knowledge. This shows that channel estimation is a key bottleneck for both methods. Nonetheless, COMMA remains to have a higher spectral efficiency for a bigger number of users. Note, that for COMMA the transmitted pilot symbols are also $q$-ary. For coded modulation with Gaussian signaling the pilot length is accordingly $Lq$ where $L$ is the number of pilot symbols for the COMMA scheme.
\section{Summary}
This work investigates uncoordinated multiple-access communication in the finite blocklength regime with an $M$-antenna receiver. The traditional approach to achieving capacity involves joint decoding of all users, but this is computationally complex. An alternative solution is to use orthogonal signaling, which reduces the complexity of MUD. Moreover, the use of chirp sequences as orthogonal basis enables low-complexity FFT-based receivers that are robust to timing and frequency offsets. This study shows that coded orthogonal modulation schemes outperform classical modulation schemes like QPSK or QAM, which we bound in terms of optimal Gaussian signaling, with linear MUD in the setting where $K_a> M$, e.g., $K_a= 500$ and $M=50$. The gap is particularly pronounced when good channel estimates are available at the receiver. To recover the transmitted symbols, a novel MMV-AMP algorithm is proposed. The study also explores the concatenation of an outer code with modified AMP detection to increase spectral efficiency for a small number of antennas $M\leq 50 -100$. Scaling laws for $M$, $K_a$, $\SNR$, and the number of orthogonal basis vectors $q$ are derived, highlighting the trade-off between receive antennas and sum-spectral efficiency. 
\printbibliography

\end{document}